\documentstyle[amsfonts,psfig,twocolumn,prb,aps]{revtex}

\newcommand{\mvec}[1]{\mbox{\boldmath $#1$\unboldmath}}
\begin{document}

\draft
\twocolumn[\hsize\textwidth\columnwidth\hsize\csname %
@twocolumnfalse\endcsname

\title{Magnetization plateaus in weakly coupled dimer spin system}

\author{Alexei K. Kolezhuk\protect\cite{perm}}
\address{Institut f\"{u}r Theoretische Physik,
Universit\"{a}t Hannover, Appelstra\ss e 2, D-30167 Hannover, Germany\\
Institute of Magnetism, National Academy of  
Sciences and Ministry of Education of Ukraine\\
36(b) Vernadskii avenue, 252142 Kiev, Ukraine}

\date{last modified September 15, 1998;  Received \today}
\maketitle

\begin{abstract}
I study a spin system consisting of strongly coupled dimers which are
in turn weakly coupled in a plane by zigzag interactions.  The model
can be viewed as the strong-coupling limit of a two-dimensional zigzag
chain structure typical, e.g., for the $(ac)$-planes of
KCuCl$_{3}$. It is shown that the magnetization curve in this model
has plateaus at ${1\over3}$ and ${2\over3}$ of the saturation
magnetization, and an additional plateau at ${1\over2}$ can appear in
a certain range of the model parameters; the critical fields are
calculated perturbatively.  It is argued that for the
three-dimensional lattice structure of the KCuCl$_{3}$ family the
plateaus at ${1\over4}$ and ${3\over4}$ of the saturation can be
favored in a similar way, which might be relevant to the recent
experiments on NH$_{4}$CuCl$_{3}$ by Shiramura {\em et al.,}
J. Phys. Soc. Jpn. {\bf 67}, 1548 (1998).

\end{abstract}

\pacs{75.10.Jm, 75.60.Ej, 75.50.Ee}

]

\section{Introduction}
\label{sec:intro}

In the last couple of years there has been a growing interest to the
phenomenon of magnetization plateaus in one-dimensional (1D) spin
systems.
\cite{Hida94,Okamoto96,Tonegawa+96,Oshikawa+97,Totsuka97,Cabra+97,%
Tonegawa+97,Narumi+97,Totsuka98,NakanoTakahashi98,SakaiTakahashi98,%
Totsuka98pre,Cabra+98pre,Mila98pre}
In a variety of models, the magnetization per site $m_{z}$ as a
function of the applied magnetic field $H$ exhibits plateaus at
certain values of $m_{z}$; at those plateaus the magnetization in
units of the saturation $M=m_{z}/S$ (here $S$ is the spin of a
magnetic ion) is ``locked'' at some rational number (at least, up to
now no indications of the possibility to have plateaus at irrational
$M$ were found). Oshikawa {\em et al.}\cite{Oshikawa+97} have shown
that in purely one-dimensional systems the allowed values of $M$, at
which the plateaus are possible, are given by the following condition
\begin{equation} 
\label{lsm} 
QS(1-M) \in \Bbb Z\,,
\end{equation}
where $Q$ is the number of magnetic ions in the magnetic elementary
cell of the ground state (which can be different from the elementary
cell prescribed by the Hamiltonian \cite{Tonegawa+97,Totsuka98}).
This condition is necessary but not sufficient; in other words, if the
plateau exists, it has to be at one of the values of $M$ determined by
(\ref{lsm}), but in principle the plateaus do not have to exist at
{\em all\/} values of $M$ which are allowed by (\ref{lsm}).  Validity
of the result (\ref{lsm}) is supported by a considerable amount of
numerical and analytical data, as well as by the recent experimental
observation \cite{Narumi+97} of the $M={1\over2}$ plateau in the $S=1$
bond-alternating chains realized in $\rm
[Ni_{2}(Medpt)_{2}(\mu$-$\rm ox)(\mu$-$\rm N_{3})]ClO_{4} \cdot 0.5H_{2}O$.

\begin{figure}
\mbox{\hspace{5mm}\psfig{figure=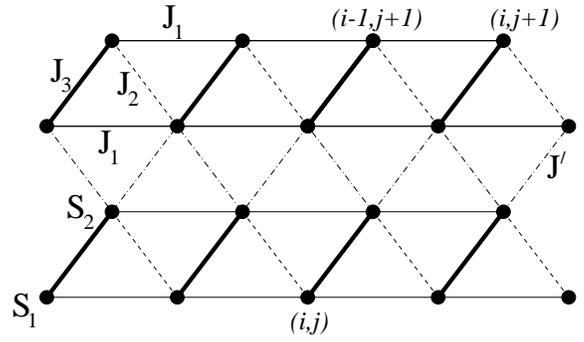,width=75mm,angle=-90}}
\vspace{3mm}
\caption{\label{fig:2dlad} Two-dimensional system of coupled zigzag
chains described by the Hamiltonian (\protect\ref{ham}), a structure
which is realized in the $(ac)$-planes of the $\rm KCuCl_{3}$ family
compounds.\protect\cite{Tanaka+96,Takatsu+97} }
\end{figure}

The present paper is motivated by the recent high-field magnetization
measurements \cite{Shiramura+98} in the double-chain $S={1\over2}$
compound $\rm NH_{4}CuCl_{3}$ which have revealed well-pronounced
plateaus at $M={1\over4}$ and $M={3\over4}$ in the temperature range
$T<1.5$~K. At room temperature, the crystal structure of this material
coincides with that of $\rm KCuCl_{3}$, and the main feature is
presence of the double chains of edge-sharing $\rm CuCl_{6}$
octahedra, separated by $\rm NH_{4}^{+}$ ions (see figs.\ 2,3 of Ref.\
\onlinecite{Tanaka+96} or fig.\ 1 of Ref.\
\onlinecite{Takatsu+97}). Double chains composed of magnetic $\rm
Cu^{2+}$ ions can be viewed as spin-$1\over2$ bond-alternating
antiferromagnetic (AF) zigzag chains with two different
nearest-neighbor exchange constants $J_{2}$, $J_{3}$ and
next-nearest-neighbor exchange $J_{1}$ (see fig.\ 1 of Ref.\
\onlinecite{Takatsu+97}, and Fig.\ \ref{fig:2dlad} of the present
paper); in what follows, the word ``chain'' is used in the meaning of
a zigzag (double) chain. If one assumes that the essential physics in
$\rm NH_{4}CuCl_{3}$ is determined by single chains, then the
appearance of plateaus at ${1\over4}$ and ${3\over4}$ of saturation
means that due to a certain rather complicated spontaneous symmetry
breaking the magnetic elementary cell contains $Q=8$ $\rm Cu^{2+}$
ions instead of $Q=2$ as suggested by the Hamiltonian symmetry; still,
then absence of plateau at $M={1\over2}$ which is characteristic for a
single chain \cite{Totsuka98} is puzzling.

Values of the exchange constants in $\rm NH_{4}CuCl_{3}$, as well as
in the other materials of the so-called $\rm KCuCl_{3}$ family,
\cite{Tanaka+96,Takatsu+97} are presently
not known. There are certain arguments \cite{NakamuraOkamoto97} based
on fitting the susceptibility data for the isostructural compound $\rm
KCuCl_{3}$ which suggest that one of the nearest-neighbor couplings
($J_{3}$) is dominating.\cite{comment1} On the other hand, neutron
scattering data for $\rm KCuCl_{3}$ strongly suggest that
interchain interactions are important. \cite{Kato+97} From the
susceptibility data,\cite{Takatsu+97,Shiramura+98} one can guess that
the intrachain couplings in $\rm NH_{4}CuCl_{3}$ are much weaker than
in $\rm KCuCl_{3}$ and thus interchain couplings should play even more
important role; this assumption also fits well to the fact that the
zero-field ground state of $\rm NH_{4}CuCl_{3}$ is magnetic at low
temperature. \cite{Shiramura+98}

In the present paper I show that interchain couplings can considerably
affect the number and positions of the plateaus. To demonstrate that,
I consider a simple model describing a system of zigzag chains coupled
in a plane; within the effective spinless
fermion model valid in the limit of weakly coupled dimers it is shown
that this model can exhibit magnetization plateaus at ${1\over3}$,
${1\over2}$, and ${2\over3}$ of the saturation. The critical fields
are calculated using the lowest-order perturbation theory around the
``atomic limit'' of the model. Physics of the plateau at $M={1\over2}$
is determined by single chains and is almost exactly the same as
studied in Ref.\ \onlinecite{Totsuka98}, while the plateaus at
$M={1\over3},{2\over3}$ appear solely due to the two-dimensional
interchain couplings. Generally speaking, the number and positions of
the ``additional'' plateaus caused by interchain interactions strongly
depend on the topology of the specific lattice.  Possible relevance of
this mechanism to the experiments on $\rm NH_{4}CuCl_{3}$ is
discussed: it is argued that the three-dimensional lattice structure
of $\rm NH_{4}CuCl_{3}$ naturally favors the appearance of plateaus
at ${1\over4}$ and ${3\over4}$ of the saturation. We also discuss
possible reasons of absence of the other plateaus in this material.

The paper is organized as follows: in Sect.\ \ref{sec:2d} the
effective model for weakly coupled dimer system is introduced and the
magnetization curve of the two-dimensional system of coupled chains is
studied, Sect.\ \ref{sec:3d} is devoted to the analysis of
characteristic features of three-dimensional coupled chain system
typical for $\rm NH_{4}CuCl_{3}$, and  Sect.\ \ref{sec:summary}
contains discussion and concluding remarks.

\section{Two-dimensional coupled chain model}
\label{sec:2d}

Consider a system of zigzag chains coupled in a plane by frustrating
zigzag interaction $J'$, as shown in Fig.\ \ref{fig:2dlad}.  Such a
structure is typical, e.g., for $(ac)$-planes of
$\rm KCuCl_{3}$. Let us further assume that the exchange
constant $J_{3}$ is antiferromagnetic (AF) and much stronger than all
the others, then the system can be viewed as a two-dimensional
arrangement of dimers weakly coupled with each other. Assume further
that the external magnetic field $\mvec{H}$ is applied in the
$z$-direction. The model is described by the Hamiltonian
\begin{eqnarray} 
\label{ham} 
\widehat{H}&=&\sum_{ij} \Big\{ J_{3}\mvec{S}_{1}^{(i,j)}\cdot
\mvec{S}_{2}^{(i,j)} + J_{2} \mvec{S}_{1}^{(i,j)}\cdot\mvec{S}_{2}^{(i-1,j)} 
\nonumber\\ 
        &+&J_{1}(\mvec{S}_{1}^{(i,j)}\cdot
\mvec{S}_{1}^{(i+1,j)}+ \mvec{S}_{2}^{(i,j)}\cdot
\mvec{S}_{2}^{(i+1,j)})\nonumber\\
&+&J'(\mvec{S}_{2}^{(i,j)}\cdot
\mvec{S}_{1}^{(i,j+1)} +\mvec{S}_{2}^{(i,j)}\cdot
\mvec{S}_{1}^{(i-1,j+1)}) \nonumber\\
&+& g\mu_{B}\mvec{H}\cdot(\mvec{S}_{1}^{(i,j)}+\mvec{S}_{2}^{(i,j)})\Big\}\,,
\end{eqnarray}
where $g$ is the Land\'e factor and $\mu_{B}$ is the Bohr magneton.
If one neglects weak interdimer interactions $J_{1}$, $J_{2}$, $J'$
completely, than the lowest states at each dimer are the singlet
$|s\rangle$ and the $S^{z}=+1$ triplet $|t_{+}\rangle$.  We will be
interested in the regime of strong fields $g\mu_{B}H\gtrsim J_{3}$,
then the other two triplet states $|t_{0}\rangle$, $|t_{-}\rangle$ are
high in energy and their contribution can be neglected, in spirit of
Refs.\ \onlinecite{Totsuka98,Totsuka98pre,Mila98pre}. In this way one
gets a reduced Hilbert space with only two degrees of freedom per
dimer, and the problem can be formulated in terms of spinless fermions
living on the asymmetric triangular lattice as shown in Fig.\ \ref{fig:2def},
with the following effective Hamiltonian $\widehat{\cal H}$:
\begin{eqnarray} 
\label{Heff} 
\widehat{{\cal H}}&=& \mu\sum_{ij} n_{ij} +U\sum_{ij} n_{ij} n_{i+1,j} +
t\sum_{ij}(c^{\dag}_{ij}c_{i+1,j} + \mbox{h.c.})\nonumber\\  
&+& U'\sum_{ij} n_{ij} (n_{i,j+1}+n_{i-1,j+1})\\
&+&t'\sum_{ij}(c^{\dag}_{ij} c_{i,j+1} + c^{\dag}_{ij} c_{i-1,j+1}+
\mbox{h.c.})\,. \nonumber
\end{eqnarray}
Here $n_{ij}\equiv c^{\dag}_{ij}c_{ij}$, and $n_{ij}=1$ corresponds to
the $|t_{+}\rangle$ state on the $(ij)$-th dimer, while absence of
particle ($n_{ij}=0$) corresponds to the singlet. The 
magnetization per site in the saturation units $M$ coincides now with
the concentration of fermions.
The effective interaction constants $U$, $U'$, hopping amplitudes $t$,
$t'$, and chemical potential $\mu$ are given by
\begin{eqnarray} 
\label{Ut} 
&& U=(2J_{1}+J_{2})/4,\quad  t=(2J_{1}-J_{2})/4
 \nonumber\\ 
&& U'=J'/4, \quad t'=-J'/4,\quad \mu=J_{3}-g\mu_{B}H\,.
\end{eqnarray}
It is useful to note the particle-hole symmetry of the effective
Hamiltonian: written in terms of hole operators, 
(\ref{Heff}) preserves its form (up to a constant shift), with the
change only in the value of the chemical potential
\begin{equation} 
\label{hole} 
\mu \leftrightarrow \mu_{\text{hole}}=-\mu-2U-4U'\,.
\end{equation}
In what follows we assume that the effective interaction constants
$U$, $U'$ are {\em positive,\/} i.e.\ the interaction is repulsive.

\subsection{The ``atomic limit''}
\label{subsec:atomic}

Consider first the ``atomic limit'' of the model (\ref{Heff}), i.e.\
put hopping $t$, $t'$ to zero. Then the problem becomes equivalent to
the generalized Ising model in magnetic field, a class of models which
is well studied in the theory of alloys (see, e.g., Ref.\
\onlinecite{Ducastelle91} and references therein). Generally, in this limit
the interaction energy per site $E_{\rm int}$ 
is a piecewise linear function of the concentration
$M$, and the curve $E_{\rm int}(M)$ has  ``kinks'' (jumps of the first
derivative) at certain values of $M$.
When the magnetic field increases, the chemical potential $\mu$ 
becomes negative, and particles start being ``pumped'' into the
system until this process is stopped by the interaction. Kinks in
$E_{\rm int}(M)$ correspond to plateaus in the 
magnetization curve $M(H)$, and the width of a plateau is proportional to the
magnitude of jump in the first derivative. The atomic limit
determines the possible number and positions of plateaus; hopping
generally tends to smear out the plateaus, and below we will study
this effect perturbatively.

\begin{figure}
\mbox{\hspace{-1mm}\psfig{figure=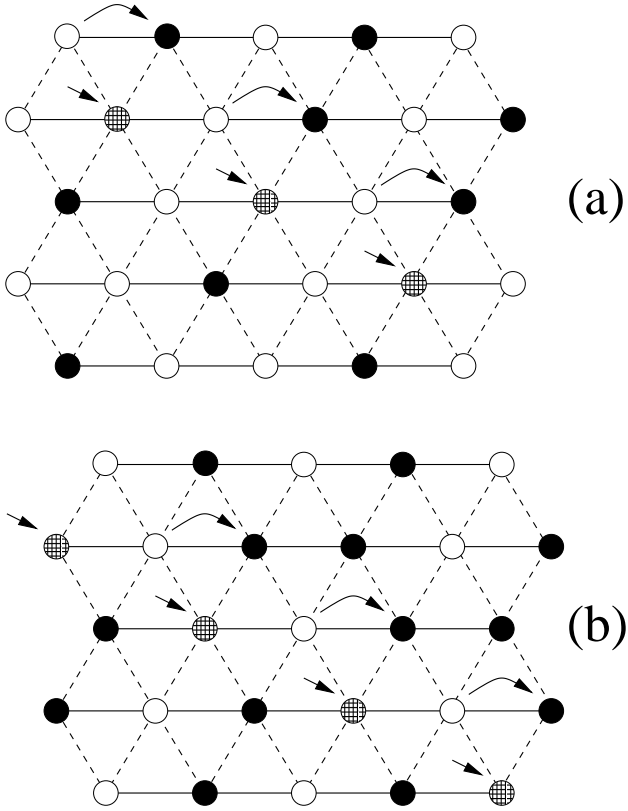,width=83mm}}
\vspace{3mm}
\caption{\label{fig:2dproc} Different processes of increasing the
concentration in the atomic limit of the model (\protect\ref{Heff}) at $U'<U$:
(a) the process leading from $M={1\over3}$ to the intermediate plateau at
$M={1\over2}$; (b) the process increasing the magnetization above
$M={1\over2}$.
Half-filled and black circles
represent respectively `new' particles inserted into the system and `old' ones
which formed the $M={1\over3}$ or $M={1\over2}$ structure.  }
\end{figure}

\begin{figure}
\mbox{\hspace{0mm}\psfig{figure=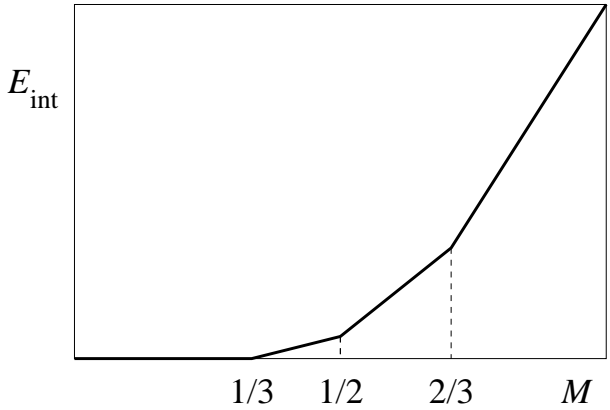,width=80mm,angle=-90}}
\vspace{3mm}
\caption{\label{fig:eint} 
A schematic look of the interaction energy per site $E_{\rm int}$ as a
function of the concentration $M$, in the ``atomic limit'' of
(\protect\ref{Heff}). 
}
\end{figure}

In the atomic limit the model (\ref{Heff}) is equivalent to the 
so-called anisotropic triangular nearest-neighbor Ising model in
external field, and its ground state is known exactly. \cite{LinWu79}
Up to $M={1\over3}$ one can insert particles into
the system without loosing any interaction energy (see Fig.\
\ref{fig:2def}). Thus, 
\begin{mathletters}
\label{Eint}
\begin{equation}
\label{Elow}
E_{\rm int}(M)=0,\quad M\leq {1\over3}\,. 
\end{equation}
Using the particle-hole symmetry, one immediately obtains from the above the
behavior of $E_{\rm int}$ above $M={2\over3}$:
\begin{equation}
\label{Ehigh}
E_{\rm int}(M)=(2U+4U')(M-{1\over2}),\quad M\geq {2\over3}\,. 
\end{equation}
In the ``intermediate'' interval of concentrations ${1\over3}< M < {2\over3}$
the situation depends on the ratio of interaction constants $U$ and $U'$. If
$U'< U$, then the lowest energy cost for inserting new particles above
$M={1\over3}$ is $3U'$ per particle, which is determined by the {\em
multiparticle} process of inserting at once a {\em row} of particles as shown
in Fig.\ \ref{fig:2dproc}a.  This process leads from the state with
$M={1\over3}$ to the state with $M={1\over2}$, and another multiparticle
process shown in Fig.\ \ref{fig:2dproc}b leads from $M={1\over2}$ to
$M={2\over3}$, taking the energy of $2U+U'$ per particle.

For $U' > U$, similar multiparticle processes exist which govern the increase
of concentration above $M={1\over3}$ and $M={1\over2}$, with the energy cost
of $3U$ and $4U'-U$ per particle, respectively (one can obtain the
corresponding pictures by rotating the particle configurations of
Fig. \ref{fig:2dproc}a,b  by $\pi/3$ but keeping the solid and dotted lines
fixed).
 
Taking all that into account, one
obtains the following result for the interaction energy as a function
of $M$ in the $[{1\over3},{2\over3}]$ interval:
\begin{equation} 
\label{Eim1}
E_{\rm int}=\cases{3U'(M-{1\over3}), &  
${1\over3}< M \leq {1\over2}$\cr
(2U+U')M-U, &
${1\over2}< M < {2\over3}$} 
\end{equation}
for $U'<U$, and
\begin{equation} 
\label{Eim2}
E_{\rm int}=\cases{3U(M-{1\over3}), &  
${1\over3}< M \leq {1\over2}$\cr
(4U'-U)(M-{1\over2})-{U\over2}, &
${1\over2}< M < {2\over3}$}
\end{equation}
for $U'>U$.
The resulting behavior of $E_{\rm int}(M)$ is schematically shown in
Fig.\ \ref{fig:eint}. Thus, generally three plateaus are possible, at
$M={1\over3}$, ${1\over2}$, and ${2\over3}$. In the limit of zero
interchain interaction $U'=0$ only the plateau at $M={1\over2}$
survives, so that one may think of this plateau as coming from single
chains, while plateaus at ${1\over3}$ and ${2\over3}$ appear due to 2D
interchain coupling. The plateau at $M={1\over2}$ disappears when 2D
interactions become isotropic (i.e., $U'=U$).
\end{mathletters}

\begin{figure}
\mbox{\hspace{5mm}\psfig{figure=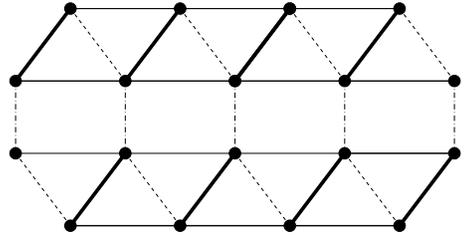,width=60mm,angle=-90}}
\vspace{3mm}
\caption{\label{fig:trellis} 
Dimers coupled in a trellis lattice. This topology allows only one plateau at
$M={1\over2}$.  }
\end{figure}

It should be remarked that the positions of those ``additional''
plateaus strongly depend on the topology of the lattice. For
instance, if one considers a trellis lattice shown in Fig.\
\ref{fig:trellis} (this type of interchain coupling is realized, e.g.,
in $\rm Sr_{n-1}Cu_{n+1}O_{2n}$ family\cite{Takano+91-92}), 
then spinless fermions live not on a
triangular, but on a square effective lattice, and the only possible
plateau would be at $M={1\over2}$.

\subsection{The effect of hopping}
\label{subsec:hopping}

Since in absence of hopping we know the solution exactly, it is
natural to try to take hopping into account perturbatively. Then one can
determine the critical fields marking the beginning and the end of
each plateau. 

To justify the use of perturbative approach with hopping term as a
perturbation, 
one has to demand that
$J_{2}$ is close to $2J_{1}$, i.e.\ that a single zigzag chain is
close to the Shastry-Sutherland line,\cite{SS81} and that the interchain
coupling is much smaller than $J_{1}$: 
\begin{equation} 
\label{justPT} 
|2J_{1}-J_{2}|\ll J_{1},\quad J'\ll J_{1}\,.
\end{equation}
Comparing (\ref{justPT}) with (\ref{Ut}), one can see that it does not make
sense to consider the case $U'>U$, and in what follows we will assume that
$U'\ll U$.

\begin{figure}
\mbox{\hspace{5mm}\psfig{figure=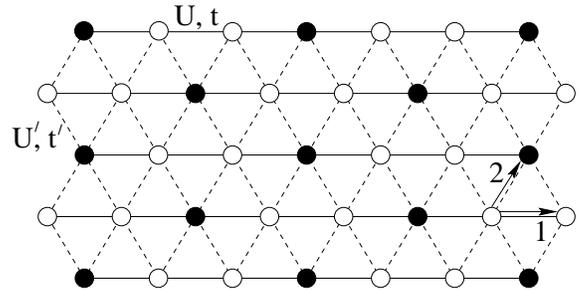,width=75mm,angle=-90}}
\vspace{3mm}
\caption{\label{fig:2def} Effective model of spinless fermions on a
triangular lattice resulting from the system shown in Fig.\
\ref{fig:2dlad}, as described by (\protect\ref{Heff}). The structure
with $M={1\over3}$ in the ``atomic limit'' is shown; open and filled
circles denote respectively empty and filled sites (i.e., dimers in
$|s\rangle$ and $|t_{+}\rangle$ states).  }
\end{figure}

Let us start with the $M={1\over3}$ structure shown in Fig.\
\ref{fig:2def}, where one third of the total number of sites $N$ is occupied
by particles. 
Increasing of the magnetization from
$M={1\over3}$ is determined by the multiparticle process of inserting
a row of $L$ new particles, $L\gg 1$ (see Fig.\
\ref{fig:2dproc}a). Then there are no first-order corrections to
$E_{N/3+L}$, and in the second order the gap
$\Delta_{c3}=(E_{N/3+L}-E_{N/3})/L$ reads as
\begin{eqnarray} 
\label{muc3-} 
\Delta_{c3}^{U'<U}&=&\mu+3U'+{4t'^{2}\over 2U-U'}+{4t^{2}\over
U} -{t^{2}\over U'}\nonumber\\
&-&{2t^{2}-12t'^{2}\over U+U'} +{5t'^{2}\over U} + O(1/L)\,.
\end{eqnarray}
Zero of the gap $\Delta_{c3}$ gives the critical field $H_{c3}$.

Similarly, for $U'<U$, increase of the magnetization from the
$M={1\over2}$ state is determined by the multiparticle process shown
in Fig.\ \ref{fig:2dproc}b, which, up to the second order, yields
the following expression for the gap $\Delta_{c5}=(E_{N/2+L}-E_{N/2})/L$:
\begin{eqnarray} 
\label{muc4-} 
\Delta_{c5}&=&\mu+2U+U'+(4t'^{2}-3t^{2})/U
+4t'^{2}/(U+U')\nonumber\\
        &+&3t^{2}/U'-6t'^{2}/(2U-U') +O(1/L)\,.    
\end{eqnarray}


To obtain the critical field $H_{c2}$, one has to consider the set of
degenerate states with one particle taken out of the structure shown
in Fig.\ \ref{fig:2def}; the degeneracy is lifted only in the second
order, giving the following dispersion
$\varepsilon_{c2}(\mvec{k})=E_{N/3-1}(\mvec{k})-E_{N/3}$ of the
$N/3-1$ excitation
\begin{eqnarray} 
\label{2d-Ec2} 
\varepsilon_{c2}(\mvec{k})&=&-\mu
+{8\over U'}[t^{2}+t'^{2}+tt'(\cos k_{1}+\cos k_{2})] \\
&+&{8t'^{2}\over
U}[1+\cos(k_{1}-k_{2})] -{3t^{2}\over U'}-{12 t'^{2}\over U+U'}\,,\nonumber
\end{eqnarray}
and the gap $\Delta_{c2}=\min \varepsilon_{c2}(\mvec{k})$ is given by
\begin{eqnarray} 
\label{muc2} 
\Delta_{c2}&=&
-\mu-3t^{2}/U'-12t'^{2}/(U+U')\nonumber\\
        &+& 8(|t|-|t'|)^{2}/U' +16t'^{2}/U, 
        \quad |t/t'| > 2U'/U, \nonumber\\
\Delta_{c2} &=&-\mu+5t^{2}/U'-12t'^{2}/(U+U')\\
        &-& 4t^{2}U/U'^{2}+8t'^{2}/U', 
\quad |t/t'| < 2U'/U\,.\nonumber
\end{eqnarray}

Finally, it is easy to calculate the value of the first critical field
$H_{c1}$ which indicates the point where the injection of triplets
starts.\cite{Mila98pre} In the vicinity of $H_{c1}$ the density of
triplets is very low, so that one can just neglect the interaction
terms in (\ref{Heff}).  The hopping part of the effective Hamiltonian
yields the dispersion $E_{1}-E_{0} \equiv \varepsilon(\mvec{k})$,
\[
\varepsilon(\mvec{k})=  \mu +2t\cos k_{1} +2t'\{\cos
k_{2}+\cos(k_{1}-k_{2})\} \,.
\]
The gap $\Delta_{c1}=\min \varepsilon(\mvec{k})$
is given by
\begin{equation} 
\label{muc1} 
\mu_{c1}=\cases{\mu-2t-(t')^{2}/t & at $t>0$\cr 
\mu+2t-4|t'| & at $t<0$}\,.
\end{equation}

All the other critical fields now can be obtained by exploiting the
particle-hole symmetry (\ref{hole}), which connects the following pairs:
\begin{eqnarray} 
\label{mu-sym} 
& \mu_{S}\leftrightarrow\mu_{c1},\quad  \mu_{c5}\leftrightarrow
\mu_{c4},& 
\nonumber\\
 &\mu_{c3}\leftrightarrow \mu_{c6},  \quad 
\mu_{c2}\leftrightarrow \mu_{c7}\,.&\nonumber
\end{eqnarray}

Zeros of the gaps given by the formulas  (\ref{muc3-}),
(\ref{muc4-}), (\ref{muc2}), (\ref{muc1}), together
with the definition of $\mu$ as stated in (\ref{Ut}), determine the
critical fields. It should be remarked that in the derivation of
(\ref{muc1}) we did not use perturbation theory, so that the expression
for $H_{c1}$ is exact within the effective model
(\ref{Heff}). It can be improved, however, by including the
processes involving the other two triplet states which were discarded
in our consideration, in spirit of the degenerate perturbation theory
(see, e.g., Ref.\ \onlinecite{tandon+98}).

\section{Three-dimensional $\bf
NH_{4}C\lowercase{u}C\lowercase{l}_{3}$-type lattice} 
\label{sec:3d}

\begin{figure}
\mbox{\psfig{figure=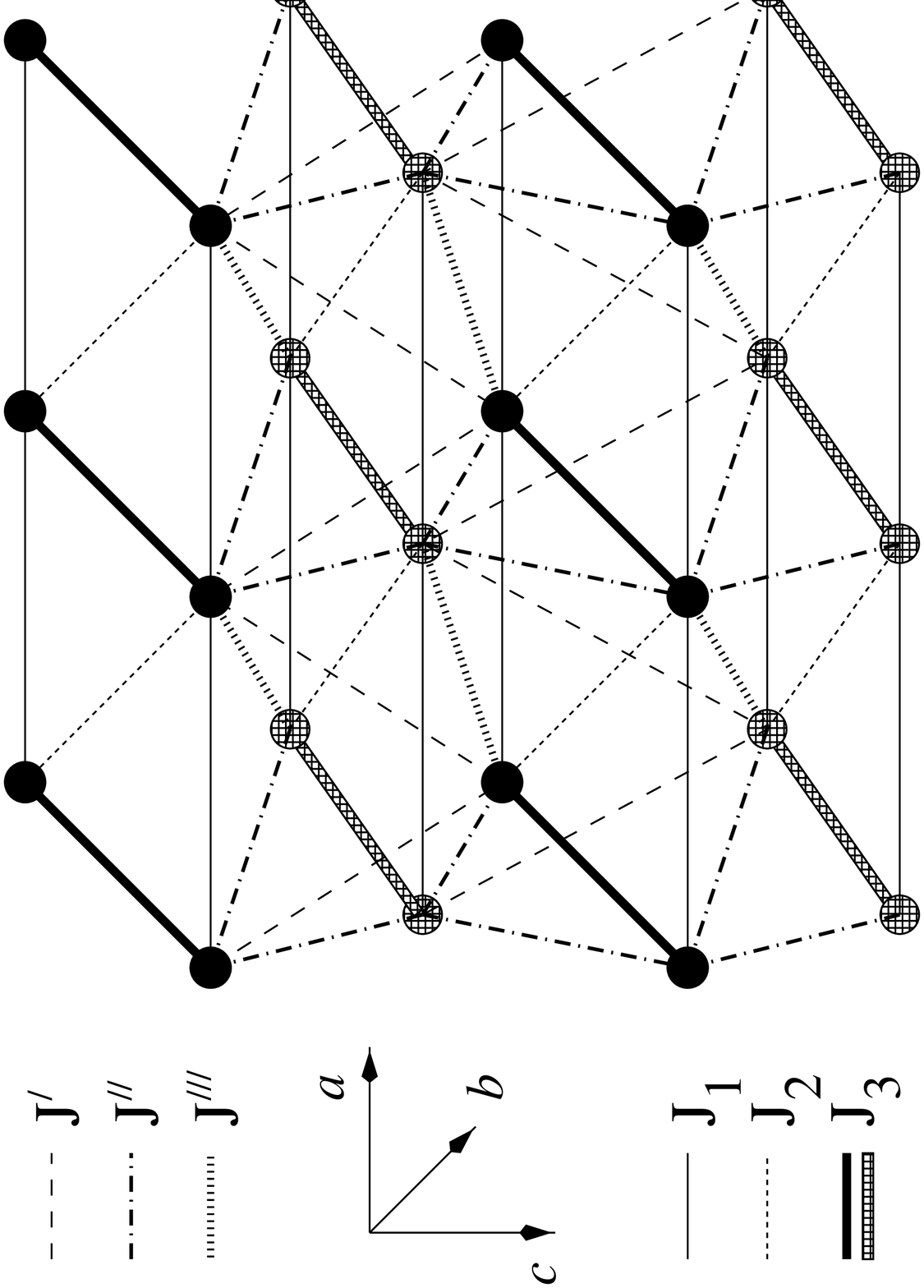,width=75mm,angle=-90}}
\vspace{3mm}
\caption{\label{fig:3dlad} 
Three-dimensionally coupled spin structure realized in $\rm KCuCl_{3}$
family. \protect\cite{Tanaka+96,Takatsu+97} Filled circles denote
$S={1\over2}$ spins of the lower layer in the $(ac)$-plane, and
half-filled circles represent spins of the upper layer.
}
\end{figure}

\begin{figure}
\mbox{\psfig{figure=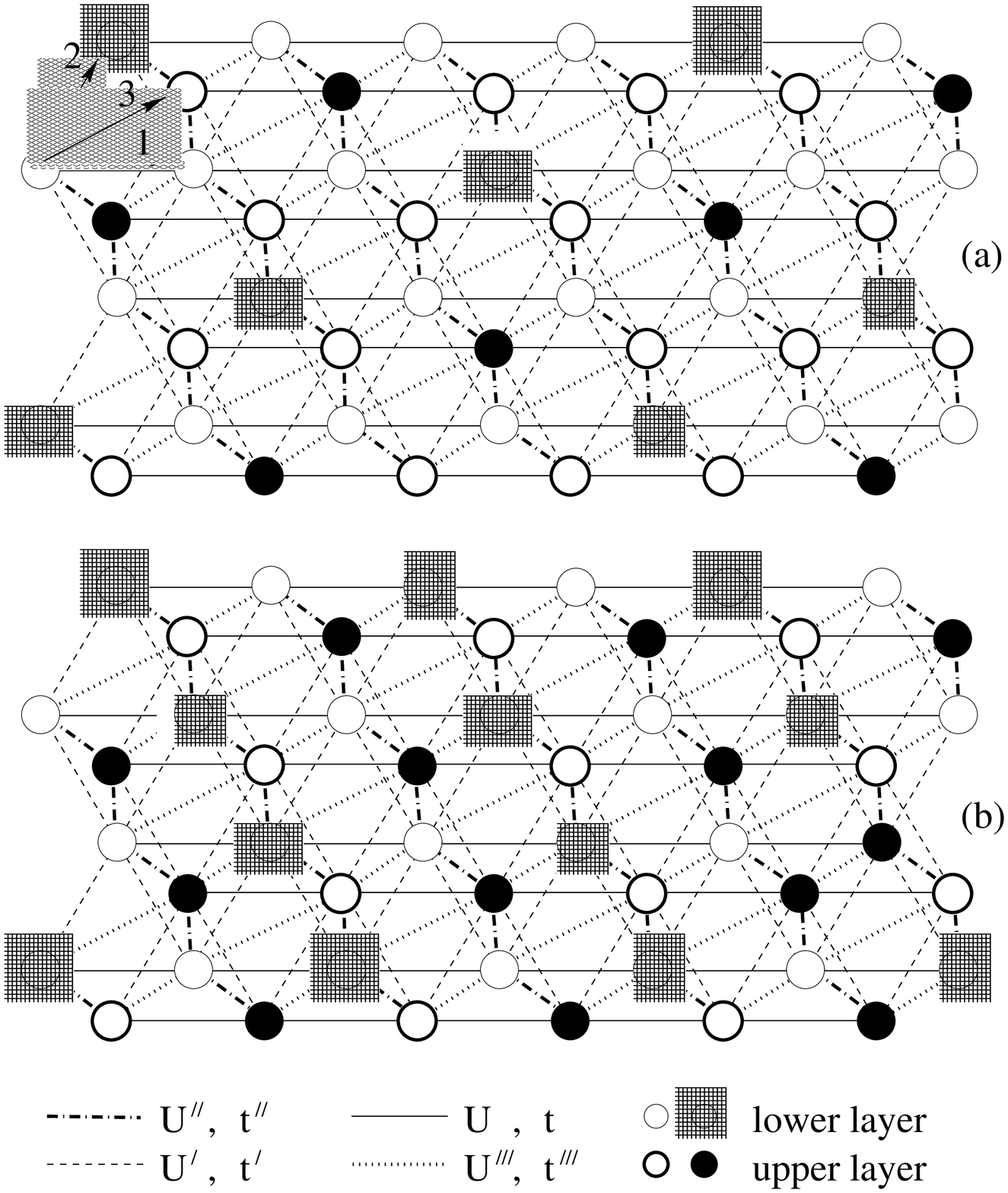,width=82mm}}
\vspace{3mm}
\caption{\label{fig:3def} Visual representation of the effective
spinless fermion model resulting from the model shown in Fig.\
\protect\ref{fig:3dlad}; the interaction constants $U$ and hopping
amplitudes $t$ are indicated for each link and are given by the
formulae (\protect\ref{Ut}), (\protect\ref{Ut+}). Filled and empty
circles represent occupied and empty sites; (a) the $M={1\over4}$ structure;
(b) the $M={1\over2}$ structure which is realized under conditions
(\protect\ref{ass}). 
}
\end{figure}

Now let us proceed to a more complicated model, which could be useful
for understanding the physics of magnetization process of $\rm
NH_{4}Cu Cl_{3}$. Consider a three-dimensional model which can be
imagined as set of two-dimensional systems (``layers'') from Section
\ref{sec:2d} stacked on top of each other, as shown schematically in
Fig.\ \ref{fig:3dlad}. Besides exchange interactions $J_{1,2,3}$, $J'$
inside a single layer, we have now two additional interlayer
interaction constants $J''$, $J'''$. This model corresponds to the
lattice structure of $\rm NH_{4}Cu Cl_{3}$,
\cite{Tanaka+96,Takatsu+97} 
with only nearest-neighbor
exchange couplings taken into account.  As before, we assume that all
interactions are antiferromagnetic, and that $J_{3}$ is much stronger
than the other exchange constants, so that the system can be viewed as
a set of weakly coupled dimers. Along the same lines of reasoning as
those used in the previous Section, one obtains the effective spinless
fermion model shown pictorially in Fig.\ \ref{fig:3def}; in addition to
(\ref{Ut}), there are now two more pairs of interaction and hopping
constants,
\begin{equation} 
\label{Ut+} 
U''={1\over2}J'',\;\; t''=0,\quad U'''={1\over4}J'''=-t'''\,.
\end{equation}
There is again a particle-hole symmetry; the relation (\ref{hole}) now
has to be modified to
\begin{equation} 
\label{3dhole} 
\mu\leftrightarrow \mu_{hole}=-\mu-2U-4U'-4U''-4U'''\,.
\end{equation}

One can find that up to the concentration $M={1\over4}$ particles can be
pumped into the system without interacting with each other. The corresponding
structure with $M={1\over4}$ is shown in Fig.\ \ref{fig:3def}a; it is easy to
see that no new particles can be inserted without interaction.
Thus, the interaction energy $E_{\rm int}$ as a function of
$M$ always has a kink at $M={1\over4}$,  
and a symmetrical kink should be present at $M={3\over4}$. Those
are exactly the values of $M$ at which the magnetization plateaus in $\rm
NH_{4}CuCl_{3}$ are observed.\cite{Shiramura+98} However, it can be shown that
in the atomic limit there are {\em always\/} other kinks at different values
of $M$ present in $E_{\rm int}(M)$.  Simply comparing the energies of
different structures and displaying them in the $E_{\rm int}(M)$ plot, one can
see that there always exists certain $M={1\over2}$ structure whose energy is
below $E_{\rm int}({1\over4})+E_{\rm int}({3\over4})$, i.e., the corresponding
point in the $E_{\rm int}(M)$ plot lies below the line connecting the
$M={1\over4}$ and $M={3\over4}$ structures; in the same way one can check that
$M={1\over3}$ and $M={2\over3}$ structures can be also favored under certain
conditions on the model parameters.

Since no other plateaus except $M={1\over4}$ and $M={3\over4}$ are
observed in $\rm NH_{4}CuCl_{3}$, this presents a problem.
Let us consider a particular regime with the exchange constants
satisfying the following inequalities:
\begin{eqnarray} 
\label{ass}
&& J''' \leq J'', \quad J' \leq 2 J''',\nonumber\\ 
&& J' \leq 2J_{1}+J_{2}+4(J''-J''')\,.
\end{eqnarray}
Then one can show that in the atomic limit the only additional plateau
structure realized in between $M={1\over4}$ and $M={3\over4}$ is the
$M={1\over2}$ structure shown in Fig.\ \ref{fig:3def}b.  Then, generally,
there exist eight critical fields, as in the 2D problem considered in the
previous section, which can be calculated perturbatively in the same way. The
perturbation theory results in three dimensions are somewhat cumbersome even
in the second order, so that they are listed in the Appendix.

The important point is that in the regime determined by (\ref{ass})
hopping reduces the width of the $M={1\over2}$ plateau already in the
first order, while corrections to the width of $M={1\over4},
{3\over4}$ plateaus start from the second order.  Thus, in the above
regime there is only one additional plateau at $M={1\over2}$ which is
smeared by hopping more effectively than the plateaus at
$M={1\over4},{3\over4}$. This could be a qualitative explanation of the
real situation in $\rm NH_{4}CuCl_{3}$, if one assumes that the
additional plateau is completely wiped out.

\section{Discussion and summary}
\label{sec:summary}

In this paper I show that in quasi-one-dimensional spin systems the
location of plateaus in the dependence of magnetization on applied
field can be considerably affected by presence of two- or
three-dimensional interchain couplings. As an example, a simple model
of two-dimensionally coupled zigzag chains is considered in a dimer
limit. It has been shown that, depending on the lattice structure and
on the interplay between the model parameters, the system can exhibit
plateaus at ${1\over3}$ and ${2\over3}$ of the saturation, in addition
to the ``usual'' plateau at one half of the saturation determined by
physics of a single chain.

We argue that this mechanism can be relevant for the recent intriguing
high-field magnetization measurements\cite{Shiramura+98} in $\rm
NH_{4}CuCl_{3}$, since the three-dimensional lattice structure of this
material naturally favors the appearance of plateaus at ${1\over4}$
and ${3\over4}$ of the saturation, i.e., exactly at those positions
where the plateaus were actually observed.

Unfortunately, the arguments presented here cannot be used for any
quantitative predictions concerning the exchange constants in $\rm
NH_{4}CuCl_{3}$, because of the following reasons: (i) zero-field
magnetic ground state of $\rm NH_{4}CuCl_{3}$ implies that the first
critical field $H_{c1}\leq 0$, and from (\ref{Ec1}) one can clearly
see that then $J_{3}$ cannot be much stronger than all the other
couplings, as we assumed; (ii) to justify considering the hopping term
as a perturbation, one would need to satisfy (\ref{justPT}), and the
condition $J_{2}\to 2J_{1}$ does not seem plausible on the basis of
the available crystal structure data. \cite{Takatsu+97}

\acknowledgments

I am deeply grateful to H.-J. Mikeska for bringing the problem of
plateaus in NH$_{4}$CuCl$_{3}$ to my attention and for subsequent
active interest in the work. It is also my pleasure to thank W. Selke
and H. Tanaka for useful comments. The hospitality of Hannover
Institute for Theoretical Physics is gratefully acknowledged.  This
work was supported by the German Ministry for Research and Technology
(BMBF) under the contract 03MI4HAN8 and by the Ukrainian Ministry of
Science (grant 2.4/27).

\appendix

\section*{Gaps in 3D model}

\begin{figure}
\mbox{\psfig{figure=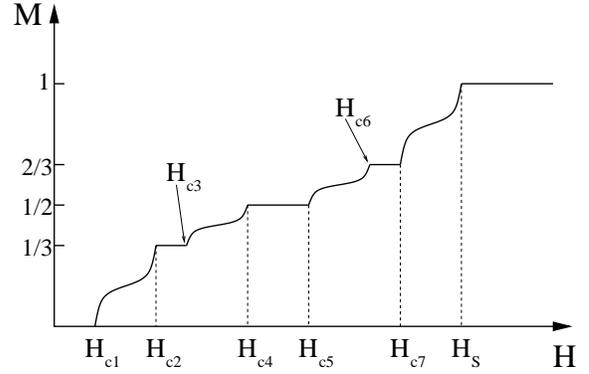,width=75mm,angle=-90}}
\vspace{3mm}
\caption{\label{fig:2dplato} 
Schematic plot of the magnetization $M$ as a function of the applied
field $H$ for the effective model (\protect\ref{Heff}).
 Solid and dashed lines show the magnetization curves
for the cases $U'\geq U$ and $U'<U$, respectively.
}
\end{figure}

In the 3D model of Sect.\ \ref{sec:3d} the magnetization curve looks similarly
to that presented in Fig.\ \ref{fig:2dplato}, except that now we have plateaus
at ${1\over4}$ and ${3\over4}$ instead ${1\over3}$ and ${2\over3}$,
respectively.  The effect of hopping can be considered perturbatively,
provided that (\ref{justPT}) is satisfied together with the additional
inequalities
\begin{equation}
\label{justPT-add}
J''\ll J_{1},\quad J'''\ll J_{1}\,.
\end{equation}
In the second order of perturbation theory, the dispersion
$\varepsilon_{c3}(\mvec{k})$ of the excitation with one extra particle
inserted above the $M={1\over4}$ state, and the dispersion
$\varepsilon_{c2}(\mvec{k})$ of the excitation resulting from taking
away one particle off the same $M={1\over4}$ state, can be represented
in the following form
\begin{eqnarray} 
\label{Ec23} 
\varepsilon_{c,i}(\mvec{k})&=&\varepsilon_{c,i}^{0} 
+A_{i}\cos(k_{1}+k_{2})+B_{i}\cos
2(k_{1}-k_{2})\\
&+&2C_{i}\cos(k_{1}-k_{2})[\cos
k_{3}+\cos(k_{1}+k_{2}-k_{3})]\,,\nonumber
\end{eqnarray}
There are no first-order corrections to the energies of  $N/4\pm1$
states. The corresponding constants for the $N/4+1$ excitation are 
given by the formulas
\begin{eqnarray} 
\label{Ec3}
\varepsilon_{c3}^{0}&=& \mu+2U'+4U''' 
+{t^{2}\over U''} +{t'^{2}\over U'''} 
+{2t^{2}\over U'+2U''} \nonumber\\
&+&{6(t^{2}+t'^{2})\over U-U'+2U''-4U'''}
+{16[t'^{2}+(t''')^{2}]\over U'+3U'''}\nonumber\\
&-&{12t'^{2}\over U'+4U'''}
+{2t'^{2}\over 2U-U'+2U''} 
 +{2t^{2}\over U+U'+2U''}
\nonumber\\
&-&{12t'^{2} \over U+2U''} -{12t^{2}\over U'+2U''} 
+{4 t'^{2}\over U+3U'' -U'''}
\\
&+&  {4t^{2}\over U'+3U''-U'''}
+{4t^{2}\over U'+2U''-U'''}
+{4(t''')^{2}\over U'+U'''}
\nonumber\\
&+&{4t'^{2}\over U+2U''-U'''}
+{2t'^{2}\over U-U'+2U''} 
-{16(t''')^{2}\over 2U'+3U'''},
\nonumber\\
A_{3}&=&{8tt'\over U-U'+2U''-4U'''} +{4(t''')^{2}\over
U'+U'''},\\
B_{3}&=&{t'^{2}\over U'''},\quad C_{3}={4t' t''' \over U'+3U'''},\nonumber
\end{eqnarray}
and for the $N/4-1$ excitation one obtains
\begin{eqnarray} 
\label{Ec2} 
\varepsilon_{c2}^{0}&=& -\mu +{8(t''')^{2}\over U'} 
+{2(t^{2}+t'^{2})\over U''} 
+{2t'^{2}\over U'''} +{2t^{2}\over U'+U''} 
\nonumber\\
&+&{2t'^{2}\over U+U''} +{16[t'^{2} +(t''')^{2}]\over U'+3U'''}
-{6t'^{2} \over U+2U''}\\
&-&{12t'^{2}\over U'+4U'''} 
-{14 (t''')^{2}\over 2U'+3U'''} -{6t^{2}\over U'+2U''}\,,\nonumber\\
A_{2}&=&{4tt'\over U''} +{4(t''')^{2}\over U'},\quad
B_{2}={t'^{2}\over U'''},\quad
C_{2}={8t' t'''\over U'+3U'''}\,.\nonumber
\end{eqnarray}

Closing of the gaps $\Delta_{c,i}=\min \varepsilon_{c,i}(\mvec{k})$
$(i=2,3)$ determines the critical fields $H_{c2}$, $H_{c3}$ marking
the beginning and the end of the $M={1\over4}$ plateau, respectively.
 Taking into account that $B_{2,3}$ and
$C_{2,3}$ are always positive [by our assumption on antiferromagnetic
character of all exchange interactions, see (\ref{Ut}), (\ref{Ut+})],
one can obtain the following formulas for the gaps:
\begin{eqnarray*} 
\Delta_{c}&=&\varepsilon_{c}^{0}+\protect\cases{
A+B-4C, & $B<C$ \cr
A-B-2C^{2}/B, & $B>C$
}\quad \mbox{at\ } A<0,
\end{eqnarray*}
and for $A>0$ the result is a bit more complicated,
\begin{eqnarray*} 
\Delta_{c}&=&\varepsilon_{c}^{0}+\protect\cases{
-A-B, &
$C<(AB)^{1/2}$ \cr 
A+B-4C, & $A<C$, $B<C$ \cr
B-A-2C^{2}/A, & $(AB)^{1/2}<C<A$\cr 
A-B-2C^{2}/B, & $(AB)^{1/2}<C<B$
}.
\end{eqnarray*}
[We have omitted  the subscripts $2,3$ above for the sake of clarity].
One can see that the  expressions for $\Delta_{c2}$, $\Delta_{c3}$
``explode'' when either of the interchain interaction constants $U'$,
$U''$, $U'''$ vanishes; this corresponds to the obvious fact that the
$M={1\over4}$ plateau ceases to exist if any of the interchain
interactions is absent. 

The energy $\varepsilon_{N/2+1}(\mvec{k})
=\varepsilon_{N/2+1}^{(1)}(\mvec{k})+\varepsilon_{N/2+1}^{(2)}(\mvec{k})$
of the excitation with one extra particle above the $M={1\over2}$
state contains both first- and second-order corrections. In the first
order approximation one obtains
\begin{eqnarray} 
\label{EN2p1-1} 
\varepsilon_{N/2+1}^{(1)}
 &=&\mu+2(U +U' +2U'')  \\
 &+&2t'\cos (k_{1}-k_{2}) +2t'''[\cos k_{3} + \cos(k_{1}+k_{3})], \nonumber
\end{eqnarray}
and the second-order contribution is
\begin{eqnarray} 
\label{EN2p1-2} 
\varepsilon_{N/2+1}^{(2)}
&=& {t^{2}(1+\cos 2k_{1})\over U''-U'''} 
+{2t'^{2}(1+\cos 2k_{2})\over U-U'+2(U''-U''')} \nonumber\\
&+&{4t^{2}+4t'^{2}
+8tt'\cos(k_{1}+k_{2})
\over U-U'+4(U''-U''')} \nonumber\\
&+&{4t^{2}+4t'^{2}+8tt'\cos(k_{1}-k_{2})\over U+4(U''-U''')} \\
&-&{22 t^{2}\over U+4(U''-U''')} -{22 t'^{2}\over 2U-U' +4(U''-U''')}
\nonumber\\ 
&+&{4t'^{2}\over 2U-U'+3(U''-U''')} 
+{4t^{2}\over U+3(U''-U''')}  \nonumber\\
&+&{4t'^{2}\over 2U-U'+3U''-4U'''}
+{4t^{2}\over U+4U''-3U'''} \nonumber\\
&+&{4t^{2}\over U+3U''-4U'''}
+{4t'^{2}\over 2U-U'+4U''-3U'''} \nonumber\\
&+&{2t^{2}\over U+U'+4(U''-U''')} 
+{2t'^{2}\over 2U+4(U''-U''')}. \nonumber
\end{eqnarray}
Zero of the corresponding gap $\Delta_{c5}=\min
\varepsilon_{N/2+1}(\mvec{k})$, together with the formulas (\ref{Ut}),
$\ref{Ut+}$, determines the critical field $H_{c5}$ marking the end
of the $M={1\over2}$ plateau.

The energy $\varepsilon_{c1}(\mvec{k})$ of the one-particle excitation
above the non-magnetic ($M=0$) state is completely determined by the
hopping part of the Hamiltonian and has the form
\begin{eqnarray} 
\label{Ec1}
\varepsilon_{c1}(\mvec{k})&=&\mu +2t\cos k_{1} 
+2t'[\cos k_{2}+\cos(k_{1}-k_{2})] \nonumber\\
&+&2t'''[\cos k_{3}+\cos(k_{1}+k_{2}-k_{3})]\,,
\end{eqnarray}
the gap in the above dispersion closes at the critical field
$H_{c1}$ above which the magnetization starts to increase from zero.

Exploiting the particle-hole symmetry (\ref{3dhole}) allows one to
connect the following pairs of critical fields:
\begin{eqnarray} 
\label{3d-mu-sym} 
& H_{S}\leftrightarrow H_{c1},\quad  H_{c5}\leftrightarrow
H_{c4},&\nonumber\\
& H_{c3}\leftrightarrow H_{c6}, \quad
 H_{c2}\leftrightarrow H_{c7} & \,.
\end{eqnarray}

Setting $H_{c4}$ equal to $H_{c5}$, we can obtain a condition for
``closing'' the $M={1\over2}$ plateau; restricting ourselves to the
first order expressions for the sake of simplicity, we get
\begin{equation} 
\label{no12} 
J_{1}+{1\over2}J_{2} +J'' \leq J'+3J'''\,.
\end{equation} 
However, it is easy to see that (\ref{no12}) is already outside the range
of validity of the perturbation theory determined by (\ref{justPT}),
(\ref{justPT-add}), so that this condition should be understood only
as a rough estimate indicating that $J'$, $J'''$ have to be
sufficiently strong to wipe out the plateau at $M={1\over2}$.

\newpage

\begin{references}
\bibitem{comment1} Those
arguments, however, have to be taken with caution since they are obtained by
fitting in a purely one-dimensional model.
\end{references}
\end{document}